\documentclass[prl,twocolumn,showpacs,preprintnumbers,amsmath,amssymb]{revtex4}

\usepackage{graphicx}
\usepackage{color}
\usepackage{amsmath}
\usepackage{epstopdf}
\usepackage{amsbsy}
\usepackage{dcolumn}
\usepackage{bm}

\begin{document}

\title{Disorder-induced freezing of dynamical spin fluctuations in underdoped cuprates}

\author{Brian M. Andersen$^1$, Siegfried Graser$^{2,3}$, and P. J. Hirschfeld$^3$}

\affiliation{
$^1$Niels Bohr Institute, University of Copenhagen, Universitetsparken 5, DK-2100 Copenhagen, Denmark\\
$^2$Theoretical Physics III, Center for Electronic Correlations and Magnetism, Institute
of Physics, University of Augsburg,
D-86135 Augsburg, Germany\\
$^3$Department of Physics, University of Florida, Gainesville, Florida 32611, USA}

\date{\today}

\begin{abstract}
We study the dynamical spin susceptibility of a correlated $d$-wave superconductor (dSC) in the presence of disorder, using an unrestricted Hartree-Fock approach.   This model provides a concrete realization of the notion that disorder slows down spin fluctuations, which eventually ``freeze out".   The evolution of disorder-induced spectral weight transfer agrees qualitatively  with experimental observations on underdoped cuprate superconductors. For sufficiently large disorder concentrations,  static spin density wave (SDW) order is created when   droplets of magnetism nucleated by impurities  overlap. We also study the disordered stripe state coexisting with a dSC and compare its magnetic fluctuation spectrum to that of the disorder-generated SDW phase.\end{abstract}


\maketitle


{\it Introduction.} Recent studies of layered high-T$_c$ superconductors have provided evidence for nematic order of the underlying electronic state\cite{ando,daou,hinkov,matsuda,lawler}. This development has reinvigorated studies of stripe phases in the cuprate materials, and shown the important role of disorder. 
 Here we focus on the magnetic excitations within a model that allows explicitly for disorder-induced pinning of incommensurate (IC) stripe spin fluctuations\cite{kivelson}. A detailed understanding of the magnetic fluctuations in high-T$_c$ materials is likely to be important for understanding the mechanism of superconductivity itself. Remarkable experimental progress in recent years has provided a converging picture of an hourglass-like dispersion of IC magnetic fluctuations which appear to be universal except for a low-energy branch which
depends, e.g, on doping and material-specific parameters\cite{tranquadareview}. All materials appear to exhibit a spin-gap beyond some critical hole concentration $x_c$. By contrast, in the underdoped regime, the fluctuations soften and eventually freeze out to form a glassy spin state. This doping dependence seems to apply to both "clean" cuprates like YBa$_2$Cu$_3$O$_{6+x}$ (YBCO) where quasi-static SDW order is found\cite{hinkov,stock06,sonier,li08,haug}, and intrinsically disordered materials like La$_{2-x}$Sr$_x$CuO$_2$ (LSCO)  where the static spin correlations are long-range, and persist for a large doping range well into the dSC dome\cite{julien}. The spin gap in the clean materials appears to persist to substantially lower doping, however.

Experimentally, it is known that an applied magnetic field or impurity substitution cause additional slowing-down of spin fluctuations\cite{hirota,katano,lake02, khaykovich02,HKimura:2003,savici,chang09}. For example, in Ref. \onlinecite{HKimura:2003} it was shown how substitution of nonmagnetic Zn ions for Cu in $15\%$ doped LSCO shifted spectral weight into the spin gap, and eventually, for enough Zn ($\sim 2\%$), generated elastic magnetic peaks in the neutron response. In YBa$_2$(Cu$_{1-y}$Zn$_y$)$_3$O$_{6.97}$ with $y=2\%$ evidence for a similar in-gap Zn-induced mode was observed by Sidis {\it et al.}\cite{sidis96} Upon increased temperature $T$ the elastic signal decreases and eventually vanishes near $T_c$ similar to an equivalent disorder signal in Zn-free LSCO.\cite{lake02,kimura99} These results are consistent with a similar slowing down and subsequent freezing seen by  $\mu$SR in underdoped cuprates\cite{muSR}, and agrees with the overall picture from NMR studies that Zn ions not only suppress dSC but also strongly enhance SDW correlations\cite{mahajan}. Recent neutron scattering off detwinned YBa$_2$Cu$_3$O$_{6.6}$ with 2\% Zn found induced short-range magnetic order and a redistribution of spectral weight from the resonance peak to uniaxial IC spin fluctuations at lower energies\cite{suchaneck}.

Theoretically, the hourglass spin dispersion has been studied within various stripe models\cite{stripe_neutron_theory,kaul}, but a quantitative description 
including both spin and charge degrees of freedom, domain formation, glassiness, and/or slow fluctuations is still lacking. Some of these issues have been addressed recently\cite{kaul,alvarez,robertson,delmaestro,andersen07,atkinson,andersen08}, but modeling the dynamics of the neutron response in the presence of disorder remains limited at present\cite{kaul,andersen07,schmid2}.

Here, we focus on the disorder-induced magnetic phase known to exist in a correlated dSC and calculate the dynamical spin susceptibility.
For a homogeneous dSC our model reduces to a system of Bogoliubov quasiparticles whose magnetic response depends crucially on the presence of a dSC gap\cite{fermiology}. In the following, we investigate the role of spatially inhomogeneous local moments induced in this quasiparticle system, i.e. our formalism includes both the spin susceptibility from the dSC condensate and the local moments. Thus it goes beyond a spin-only approach by including also the charge carriers important for these metallic systems. This is consistent with recent neutron experiments providing evidence for a two-component scenario where intertwined regions of itinerant carriers and local moments coexist\cite{kofu,chang07}. A main goal of the calculation is to elucidate the spin fluctuation spectrum from an SDW droplet phase and compare it to the disordered stripe phase.

{\it Model.} The Hamiltonian is given by
\begin{eqnarray}\nonumber\label{Hamiltonian}
\hat{H} &\!\!\!\!=\!\!\!\!& -\!\!\sum_{ij\sigma}t_{ij}\hat{c}_{i\sigma}^{\dagger}\hat{c}_{j\sigma}\!\!+\!\!\sum_{i\sigma}(V_{i}\!\!-\!\!\mu)\hat{n}_{i\sigma} +U\sum_{i\sigma}\frac{n_{i}-\sigma m_{i}}{2}\hat{n}_{i\sigma}\\
 &+&\sum_{i\delta}\left( \Delta_{\delta i}c_{i\uparrow}^{\dagger}c_{i+\delta\downarrow}^{\dagger}+\Delta_{\delta i}^{*}c_{i+\delta\downarrow}c_{i\uparrow}\right),
\end{eqnarray}
where $\hat{c}_{i\sigma}^\dagger$ creates an electron on site $i$
with spin $\sigma$, and $t_{ij}=\{t,t'\}$ denote the two
nearest neighbor (NN) hopping integrals, $V_i$ is an impurity potential from a
set of $N$ point-like
scatterers of strength $V=100t$, $\mu$ is the chemical potential
and $\Delta_{ij}$ is the $d$-wave pairing potential between sites
$i$ and $j$. The amplitude of $\Delta_{ij}$ is set by the dSC
coupling constant $g$. Below we fix the parameters
$t'=-0.35t$, g=0.3t leading to pairing amplitudes $\Delta\sim 0.1t$, and adjust $\mu$ to give a hole doping $x=1-n\simeq 0.13$. We solve
Eq.(\ref{Hamiltonian}) self-consistently on unrestricted  $24\times 24$ lattices by diagonalizing the
associated Bogoliubov-de Gennes (BdG) equations at $T=0$\cite{JWHarter:2006}.

The model given by Eq.(\ref{Hamiltonian}) has been used
extensively to study the competition between bulk dSC and SDW phases, the origin of
field-induced magnetization, and moment formation around nonmagnetic impurities\cite{alloul09}. In
the case of many impurities, Eq.(\ref{Hamiltonian}) was used to model static disorder-induced magnetic droplets\cite{alvarez,atkinson,andersen07,andersen08,schmid2}, and to explain how these may increase in volume fraction when moving to lower doping levels and eventually form a SDW long-range ordered phase.
More recently Eq.(\ref{Hamiltonian}) extended to the vortex state was used to obtain a semi-quantitative description of the $T$-dependence of the elastic neutron response in underdoped LSCO\cite{schmid2}.

The transverse bare spin susceptibility $\chi_0^{xx}(\vec{r}_{i},\vec{r}_{j},\omega)=-i\int_{0}^{\infty}dt\, e^{i\omega t}\left\langle \left[\sigma_{i}^{x}(t),\sigma_{j}^{x}(0)\right]\right\rangle$, can be expressed in terms of the BdG eigenvalues $E_n$ and eigenvectors $u_n, v_n$ as
\begin{eqnarray}\nonumber
\chi_0^{xx}(\vec{r}_{i},\vec{r}_{j},\omega) & = & \sum_{m,n} f(u,v)\frac{f(E_{m})+f(E_{n})-1}{\omega+E_{m}+E_{n}+i\Gamma}\\
 & + & \sum_{m,n} g(u,v)\frac{f(E_{m})+f(E_{n})-1}{\omega-E_{m}-E_{n}+i\Gamma},
\end{eqnarray}
\begin{eqnarray}
f(u,v) & = & u_{m,i}^{*}v_{n,i}^{*}\left(u_{m,j}v_{n,j}-u_{n,j}v_{m,j}\right),\\
g(u,v) & = & v_{m,i}u_{n,i}\left(u_{m,j}^{*}v_{n,j}^{*}-u_{n,j}^{*}v_{m,j}^{*}\right).
\end{eqnarray}

Including the electronic interactions within RPA we find for the full susceptibility
\begin{equation}
\chi^{xx}(\vec{r}_i,\vec{r}_j,\omega)\!=\!\sum_{\vec{r}_l}\left[1-U\chi^{xx}_{0}(\omega)\right]_{\vec{r}_i,\vec{r}_l}^{-1}\chi^{xx}_{0}(\vec{r}_l,\vec{r}_j,\omega).
\end{equation}
%
Fourier transforming with respect to the relative coordinate $\vec{r}=\vec{r}_i-\vec{r}_j$ defines the spatially resolved momentum-dependent
susceptibility $\chi(\vec{q},\vec{R},\omega)=\sum_{\vec{r}}e^{i\vec{q}\cdot\vec{r}}\chi(\vec{R},\vec{r},\omega)$. Averaging over the center of mass coordinate $\vec{R}=(\vec{r}_i+\vec{r}_j)/2$, this expression gives the susceptibility $\chi(q,\omega)$ relevant for comparison with neutron measurements.

\begin{figure}[b]
\begin{minipage}{.49\columnwidth}
\includegraphics[clip=true,height=0.8\columnwidth,width=0.98\columnwidth]{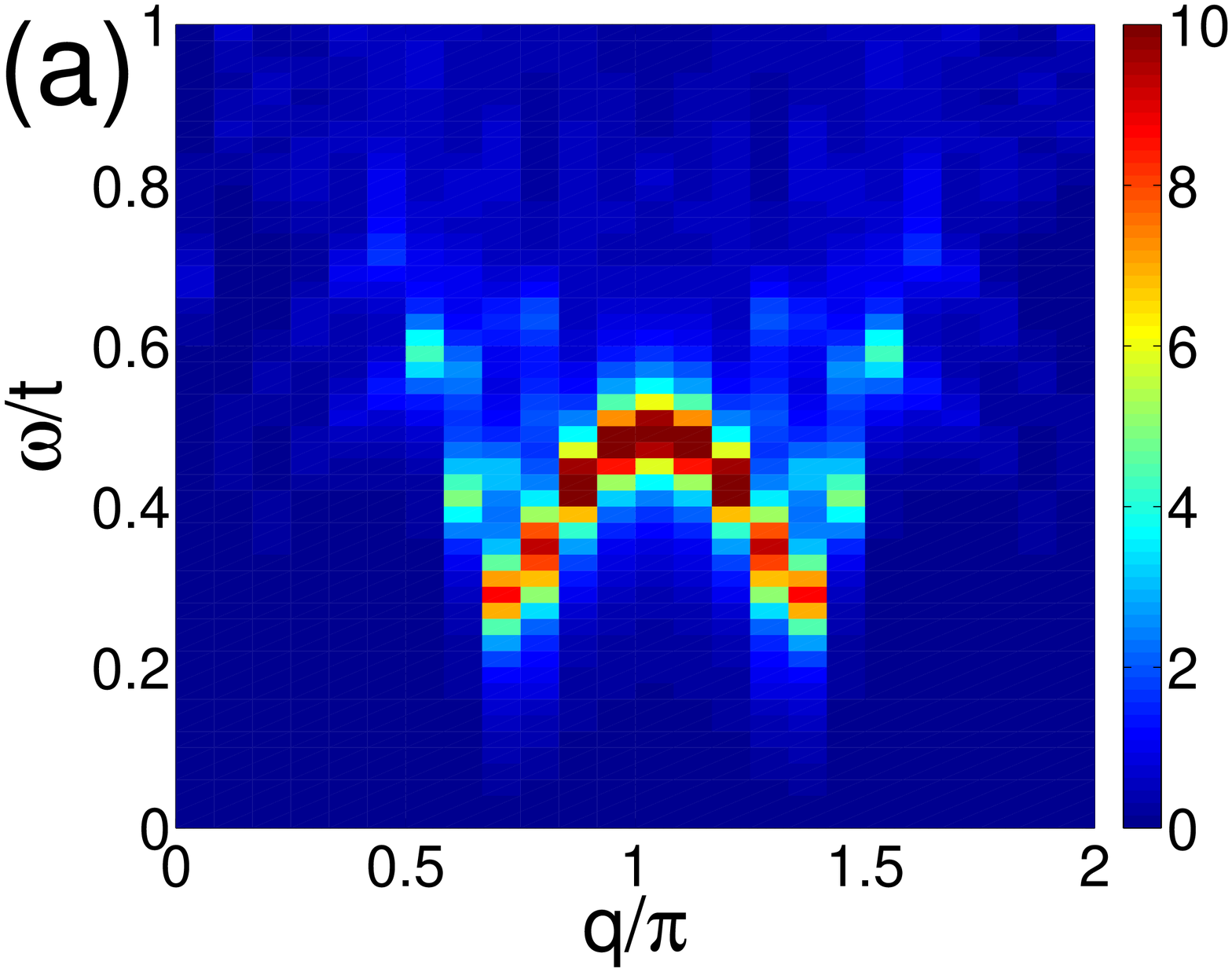}
\end{minipage}
\begin{minipage}{.49\columnwidth}
\includegraphics[clip=true,height=0.8\columnwidth,width=0.98\columnwidth]{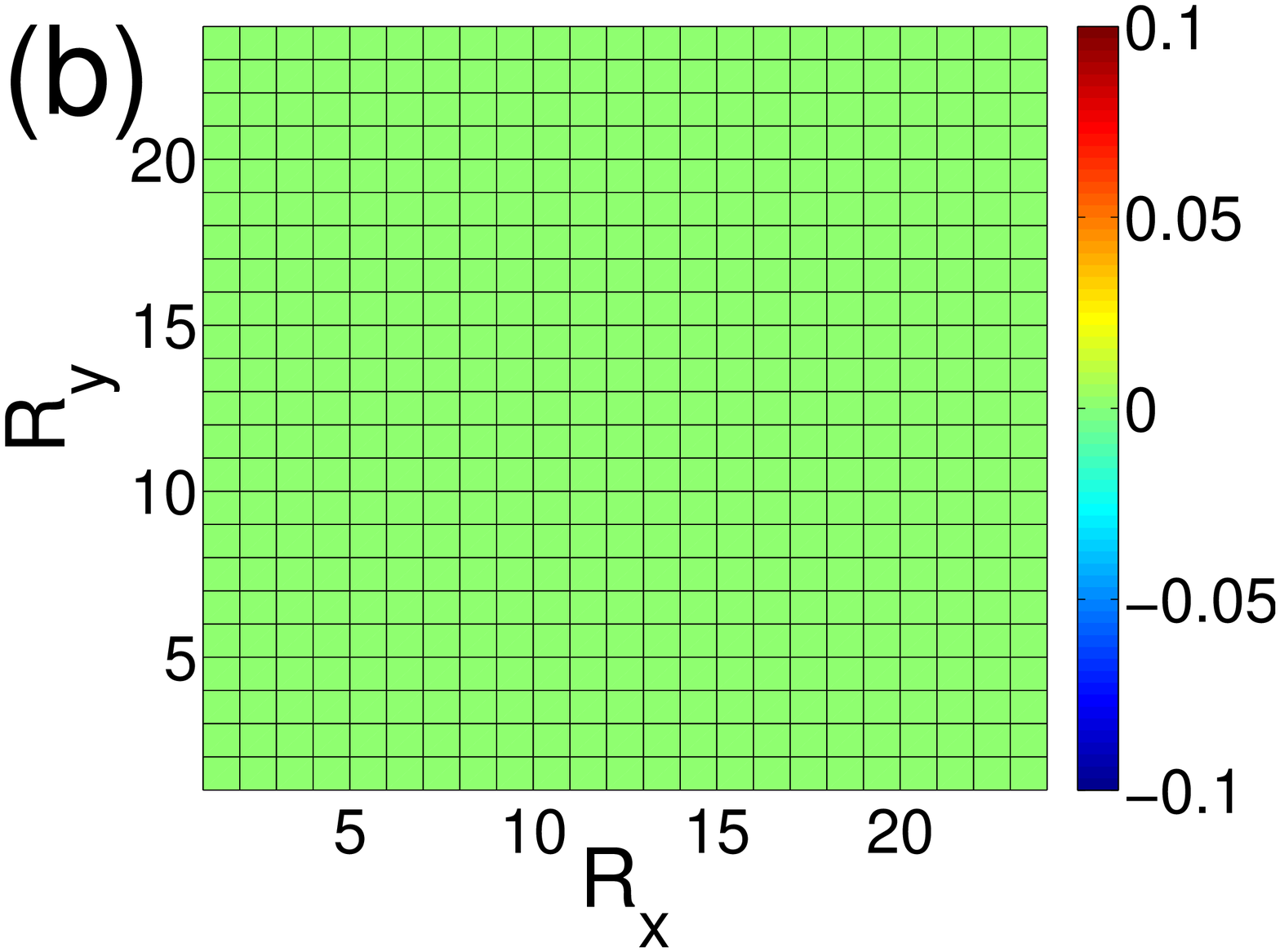}
\end{minipage}
\\
\begin{minipage}{.49\columnwidth}
\includegraphics[clip=true,height=0.8\columnwidth,width=0.98\columnwidth]{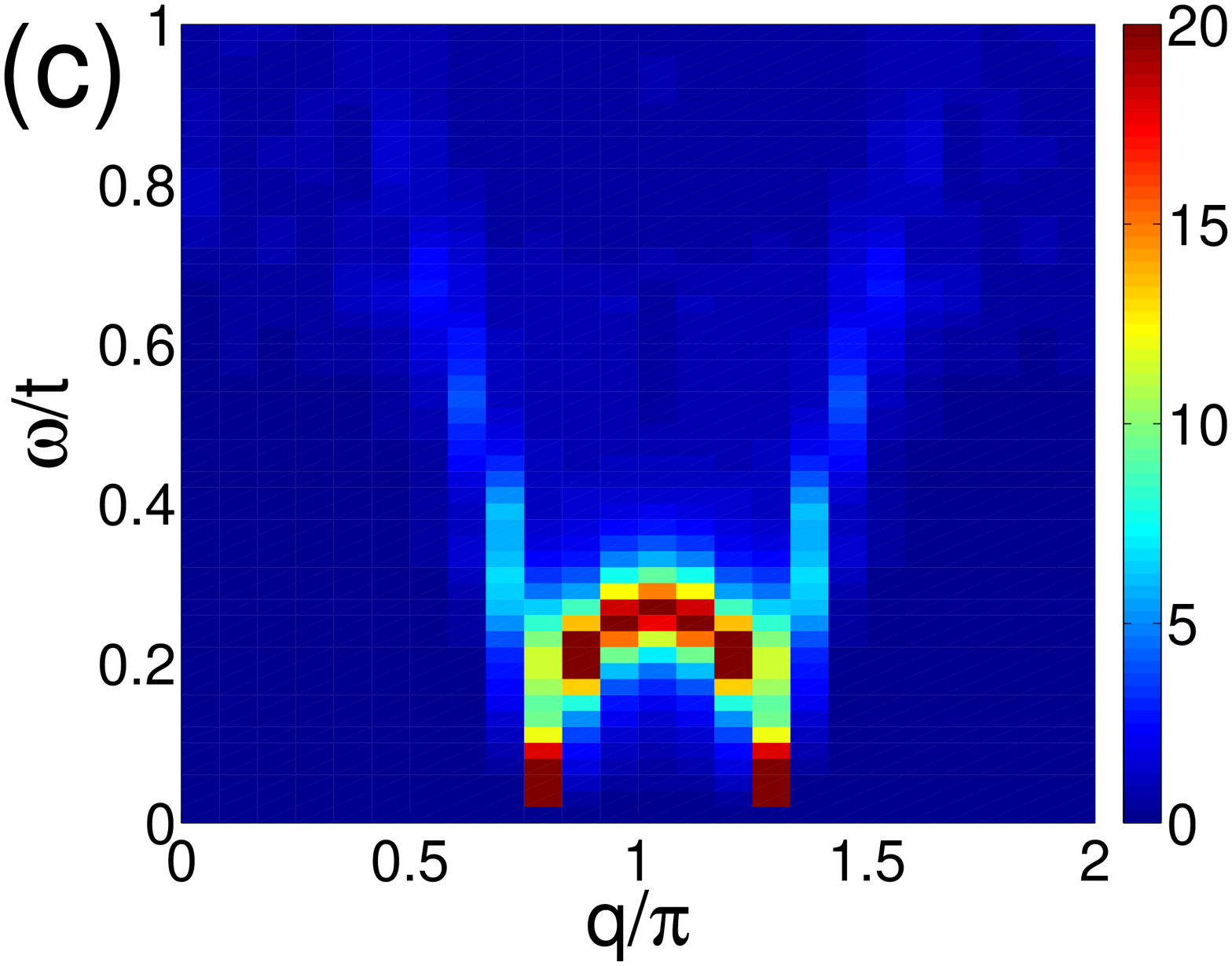}
\end{minipage}
\begin{minipage}{.49\columnwidth}
\includegraphics[clip=true,height=0.8\columnwidth,width=0.98\columnwidth]{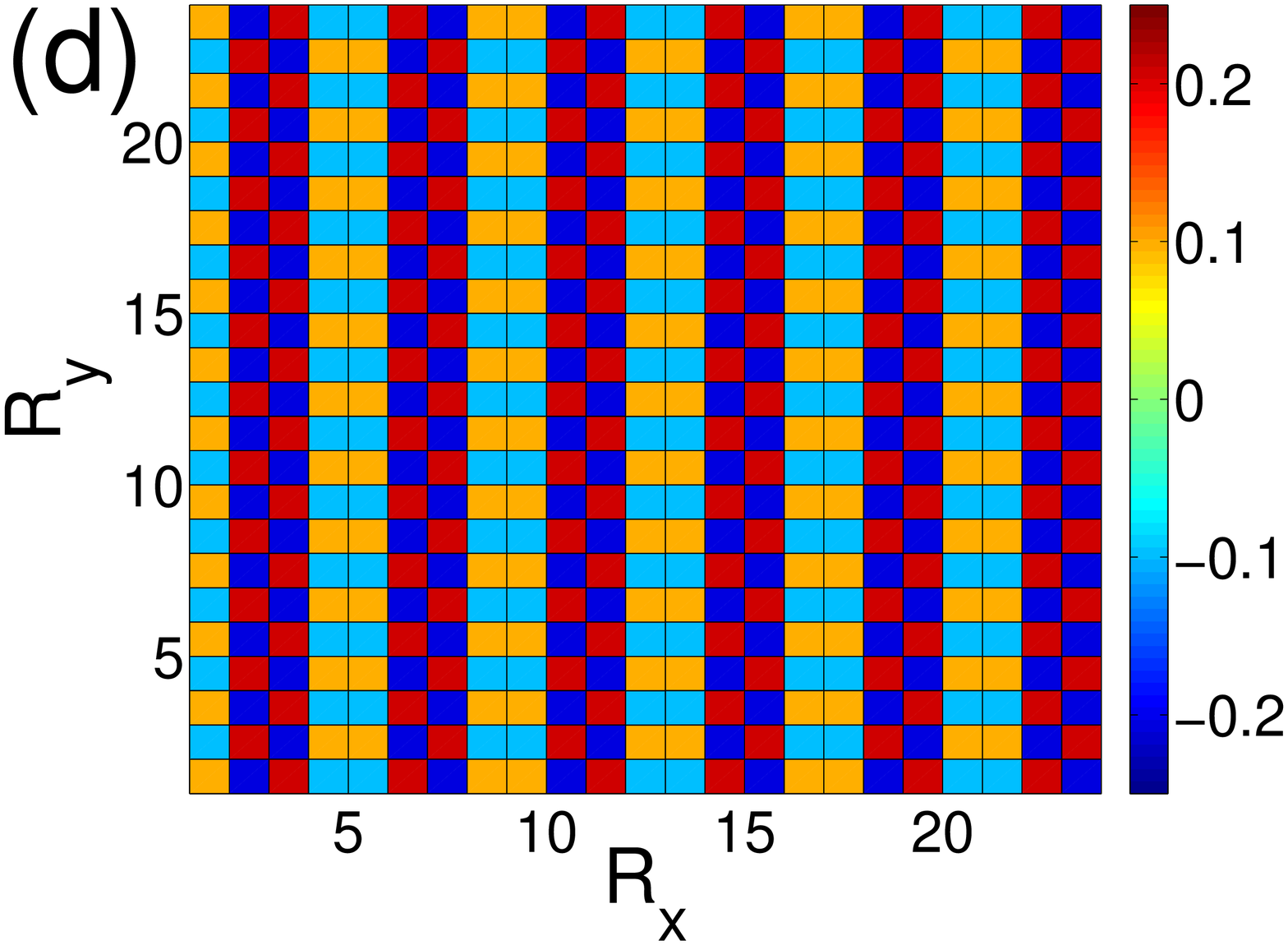}
\end{minipage}
\caption{Intensity plot of $\chi(q,\omega)$ along a cut $\vec{q}=(q,\pi)$ (a,c) and magnetization (b,d) in the clean system with (without) stripes shown in (a,b) [$U=2.6t$] ((c,d)  [$U=2.9t$]). The susceptibility in (c) is the sum of
$\chi(q,\pi,\omega)$ and $\chi(\pi,q,\omega)$ representing domain averaging.} \label{fig1}
\end{figure}


\begin{figure}[b]
\begin{minipage}{.49\columnwidth}
\includegraphics[clip=true,height=0.8\columnwidth,width=0.98\columnwidth]{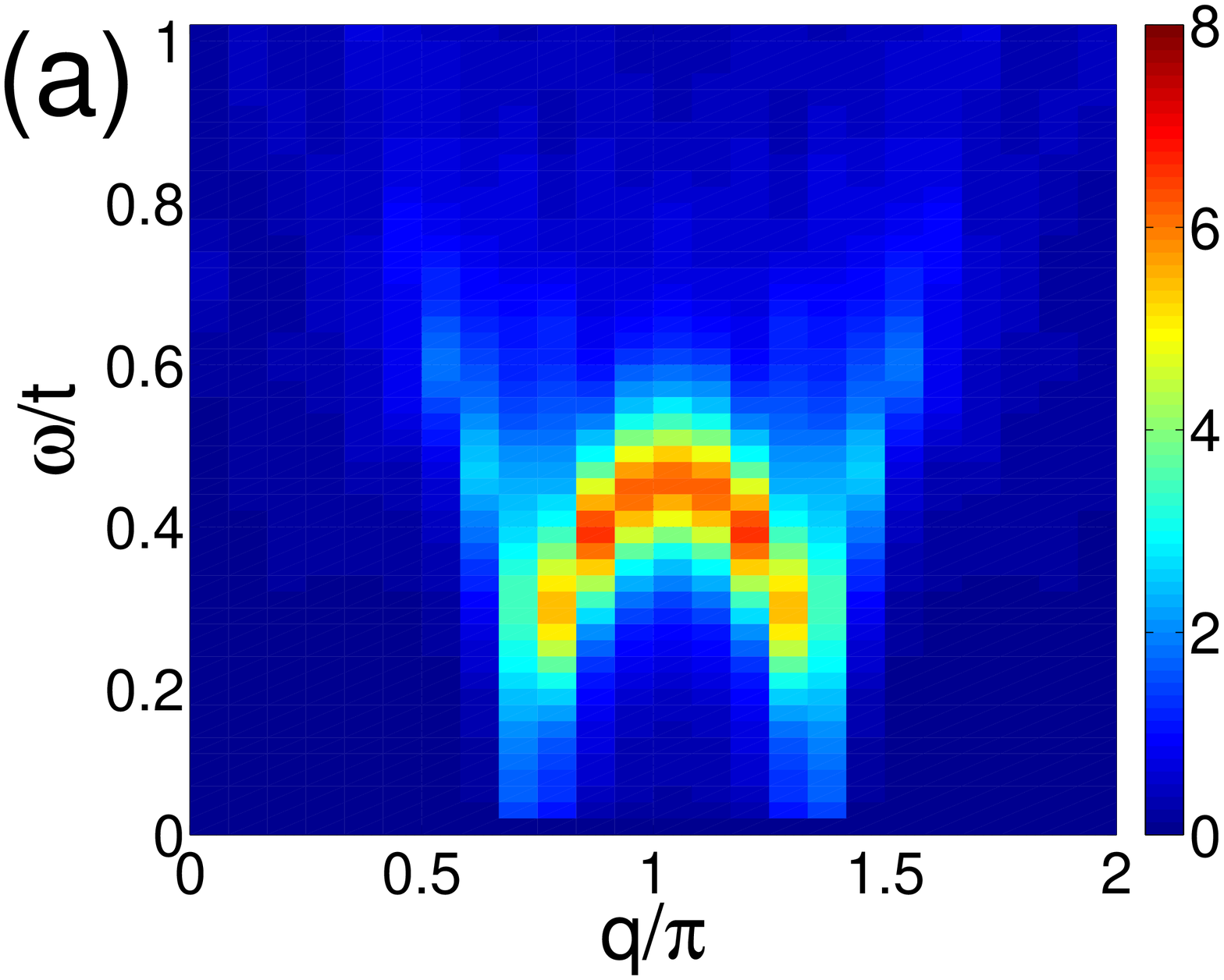}
\end{minipage}
\begin{minipage}{.49\columnwidth}
\includegraphics[clip=true,height=0.8\columnwidth,width=0.98\columnwidth]{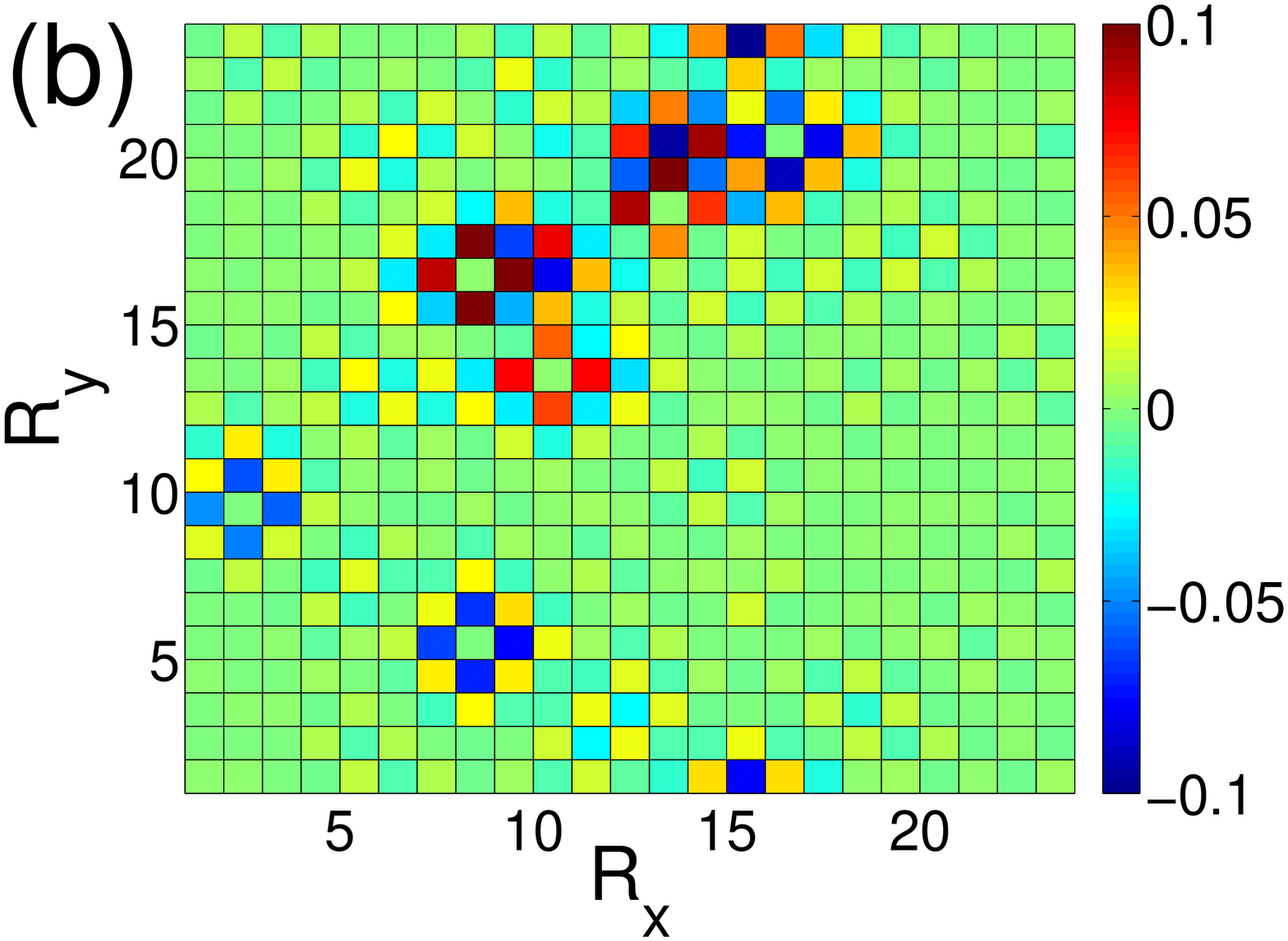}
\end{minipage}
\\
\begin{minipage}{.49\columnwidth}
\includegraphics[clip=true,height=0.8\columnwidth,width=0.98\columnwidth]{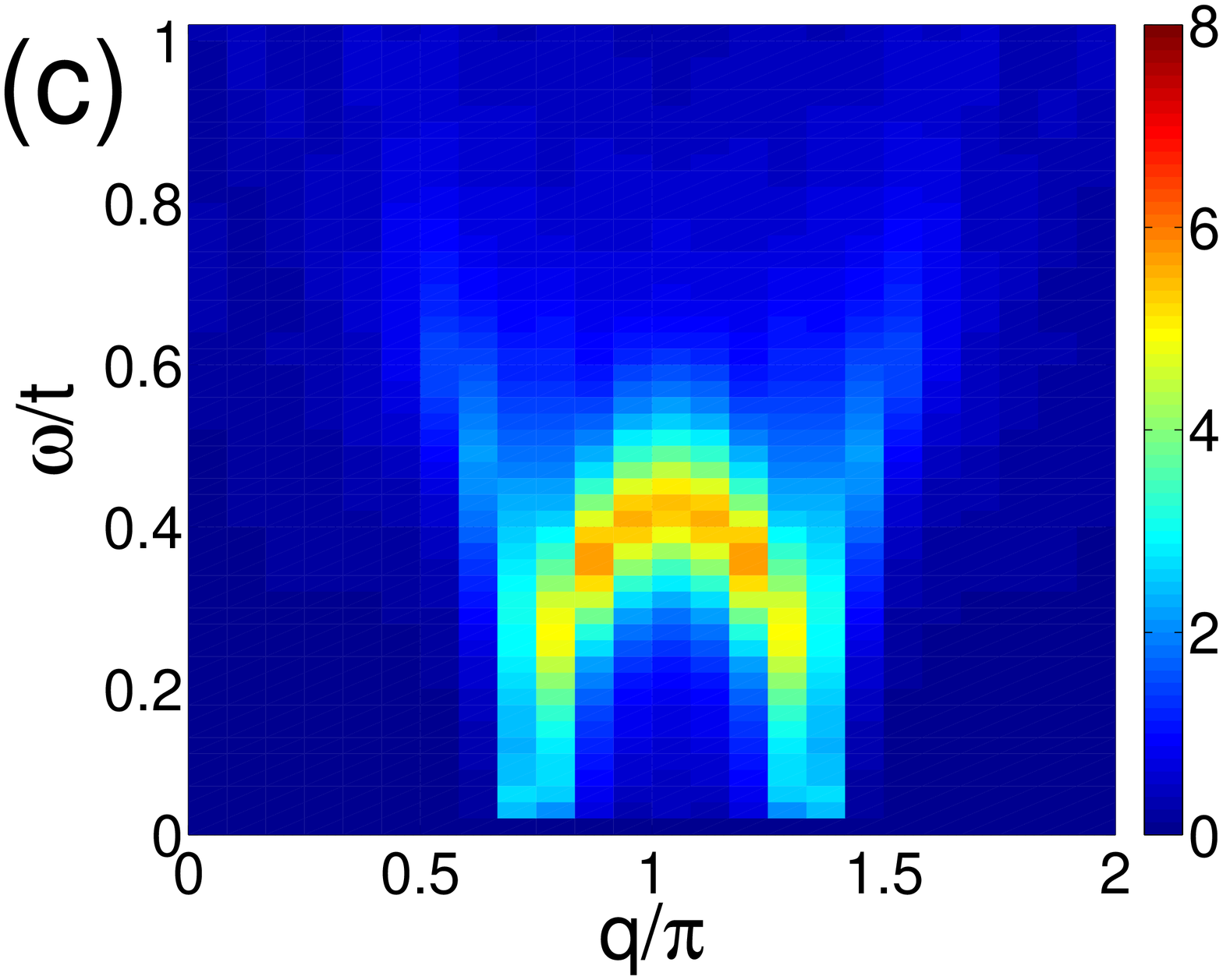}
\end{minipage}
\begin{minipage}{.49\columnwidth}
\includegraphics[clip=true,height=0.8\columnwidth,width=0.98\columnwidth]{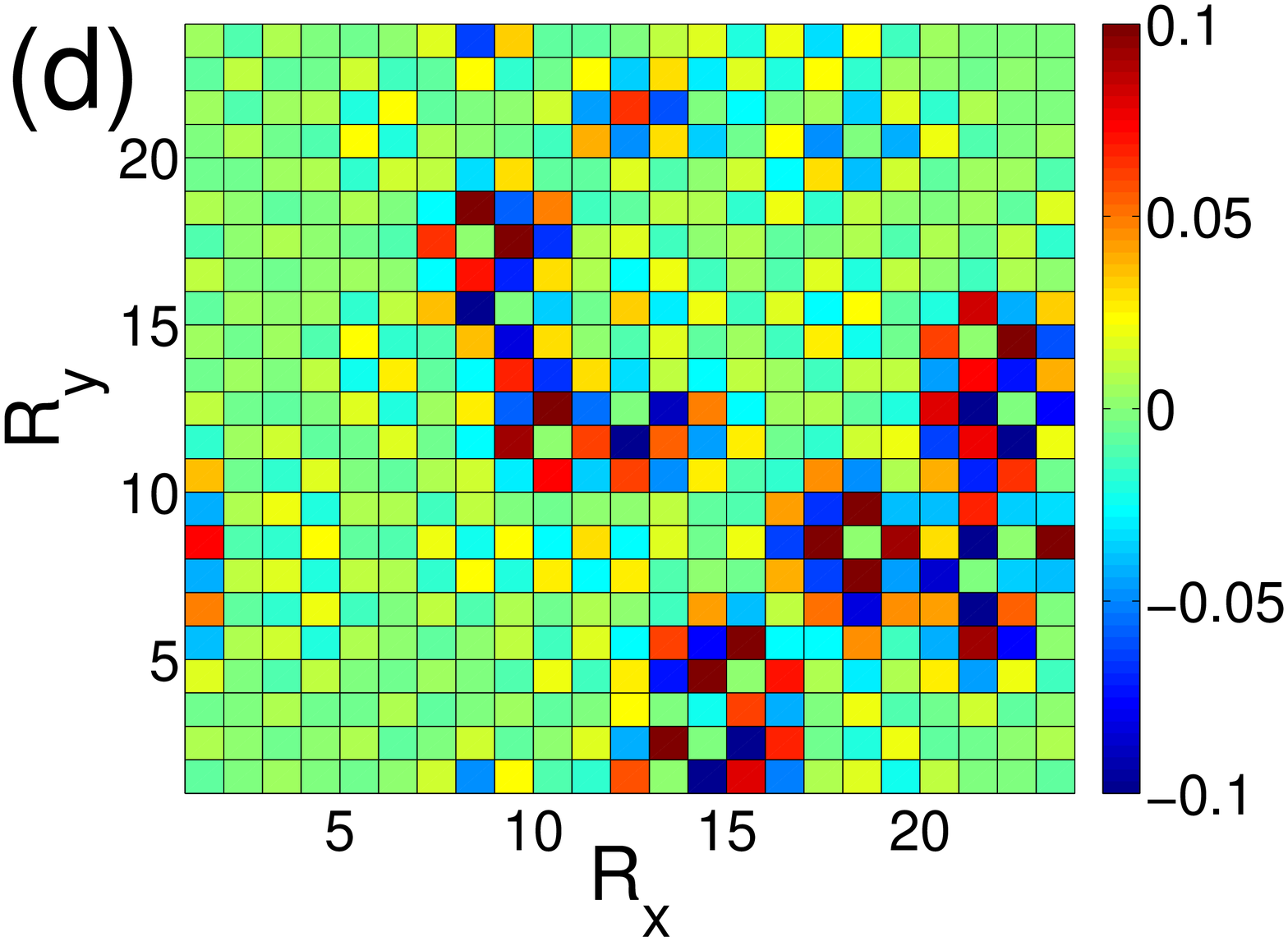}
\end{minipage}
\\
\begin{minipage}{.49\columnwidth}
\includegraphics[clip=true,height=0.8\columnwidth,width=0.98\columnwidth]{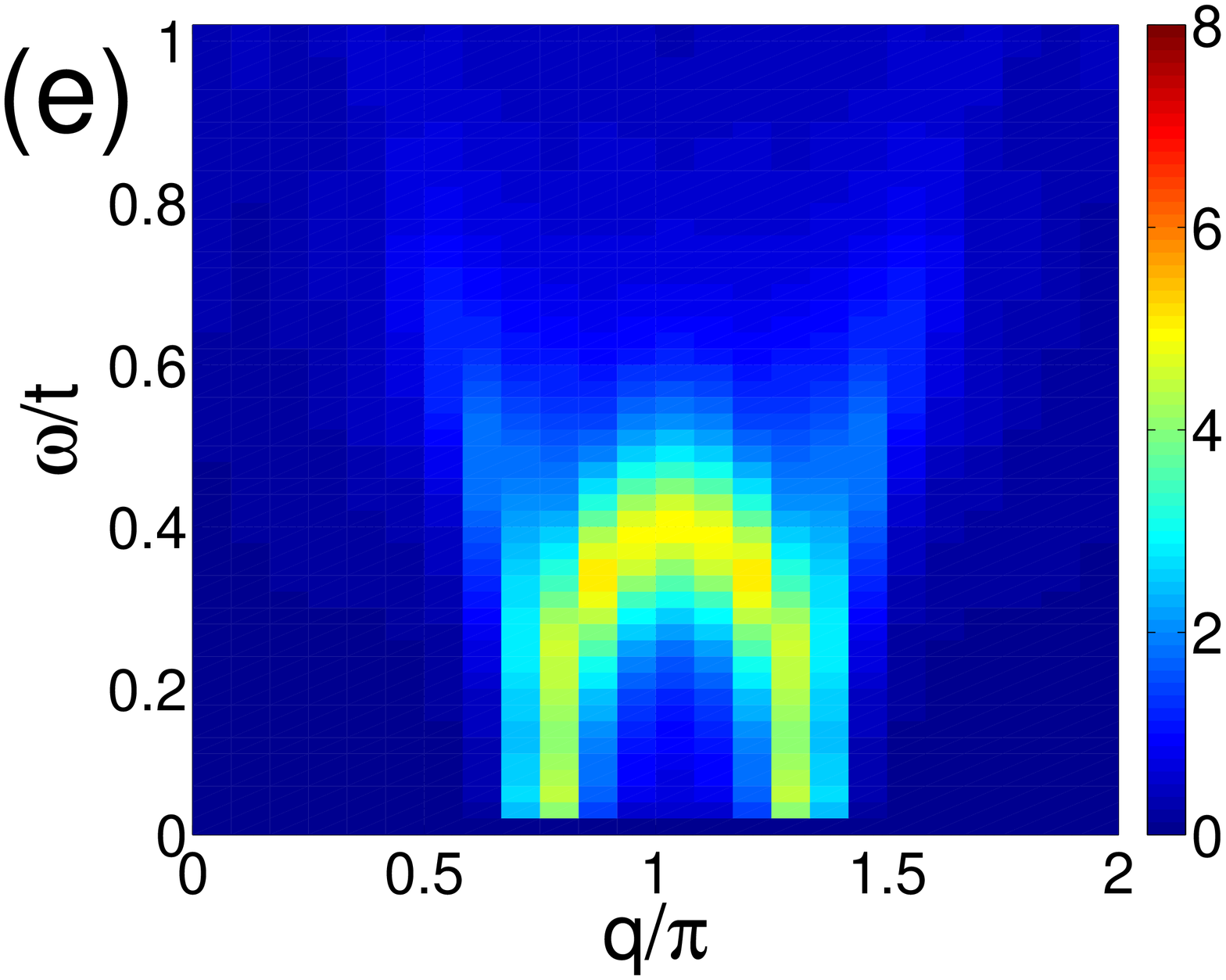}
\end{minipage}
\begin{minipage}{.49\columnwidth}
\includegraphics[clip=true,height=0.8\columnwidth,width=0.98\columnwidth]{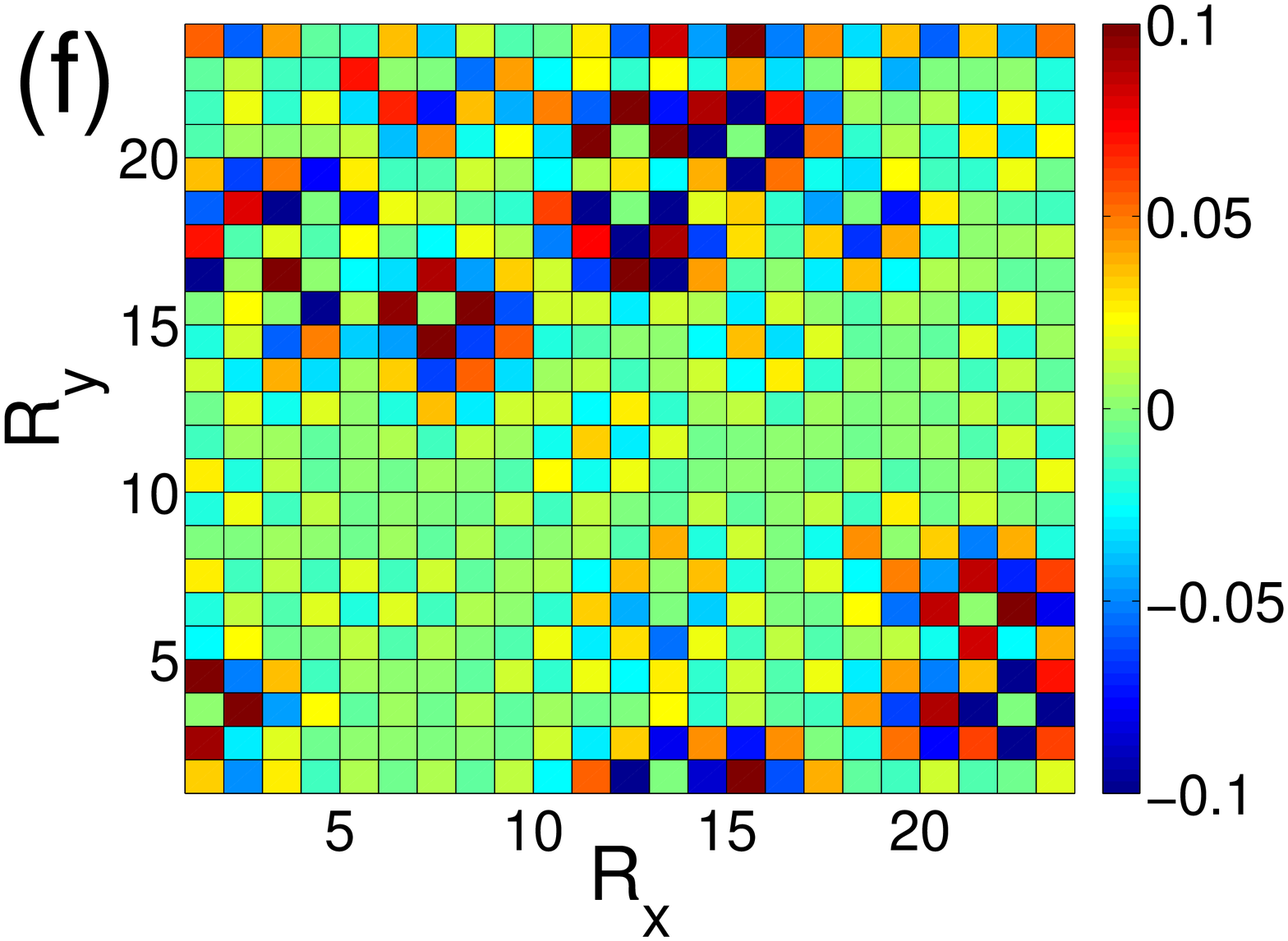}
\end{minipage}
\caption{(a,c,e) Same as in Fig. \ref{fig1}(a) with the addition of disorder. (b,d,f) show the associated real-space static magnetization. Parameters are: $U=2.6t$ and $n_i\simeq 1\%$ (a,b), $n_i\simeq2\%$ (c,d), and $n_i\simeq3\%$ (e,f). The results in the left panels are averaged over ten random disorder configurations.} \label{fig2}
\end{figure}

{\it Results.}
The Hamiltonian (\ref{Hamiltonian}) supports both a correlation- and disorder-induced SDW phase. Specifically, in the clean case above a critical repulsion $U_{c2}\simeq 2.75t$ a global stripe phase is the favorable state. Below $U_{c2}$ the ground state is a homogeneous dSC but nonmagnetic disorder may locally induce moments if $U>U_{c1}$ where $U_{c1}$ is another critical interaction strength\cite{andersen07,JWHarter:2006}. In Fig. \ref{fig1} we show the magnetization and spin susceptibility in the clean system for $U<U_{c2}$ [Fig. \ref{fig1}(a,b)] and $U>U_{c2}$ [Fig. \ref{fig1}(c,d)]. Figure \ref{fig1}(a,b) is a homogeneous dSC whereas the stripe phase in Fig. \ref{fig1}(c,d) coexists with a stripe-modulated dSC order parameter. The spin susceptibility in Fig. \ref{fig1}(a) is dominated by a spin-1 in-gap resonance mode whose very existence depends crucially on the presence of a dSC gap\cite{fermiology}. By contrast, the spin-wave branches in the ordered case in  Fig. \ref{fig1}(c) are independent of the presence of a coexisting dSC condensate.

\begin{figure}[t]
\begin{minipage}{.49\columnwidth}
\includegraphics[clip=true,height=0.8\columnwidth,width=0.98\columnwidth]{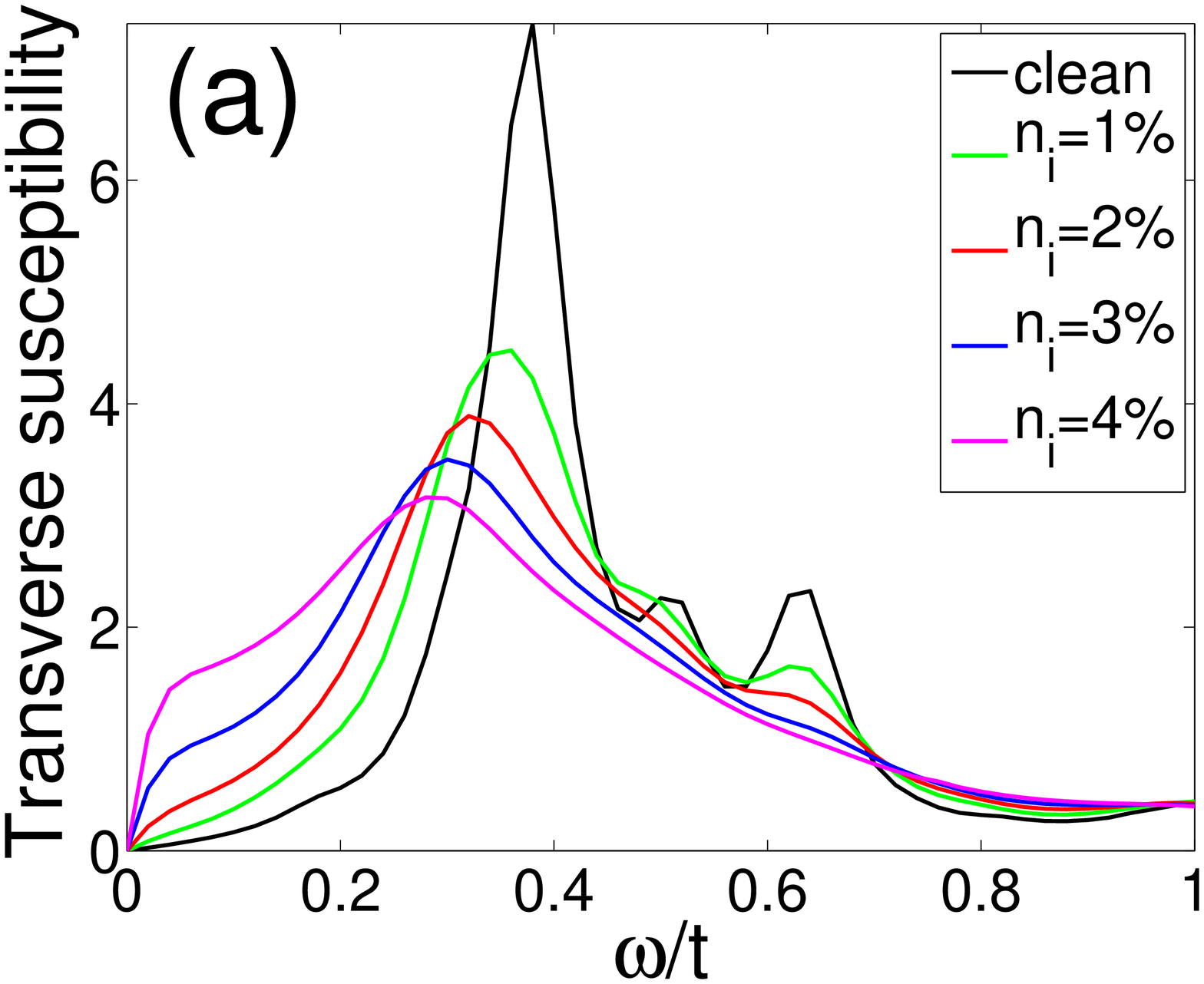}
\end{minipage}
\begin{minipage}{.49\columnwidth}
\includegraphics[clip=true,height=0.8\columnwidth,width=0.98\columnwidth]{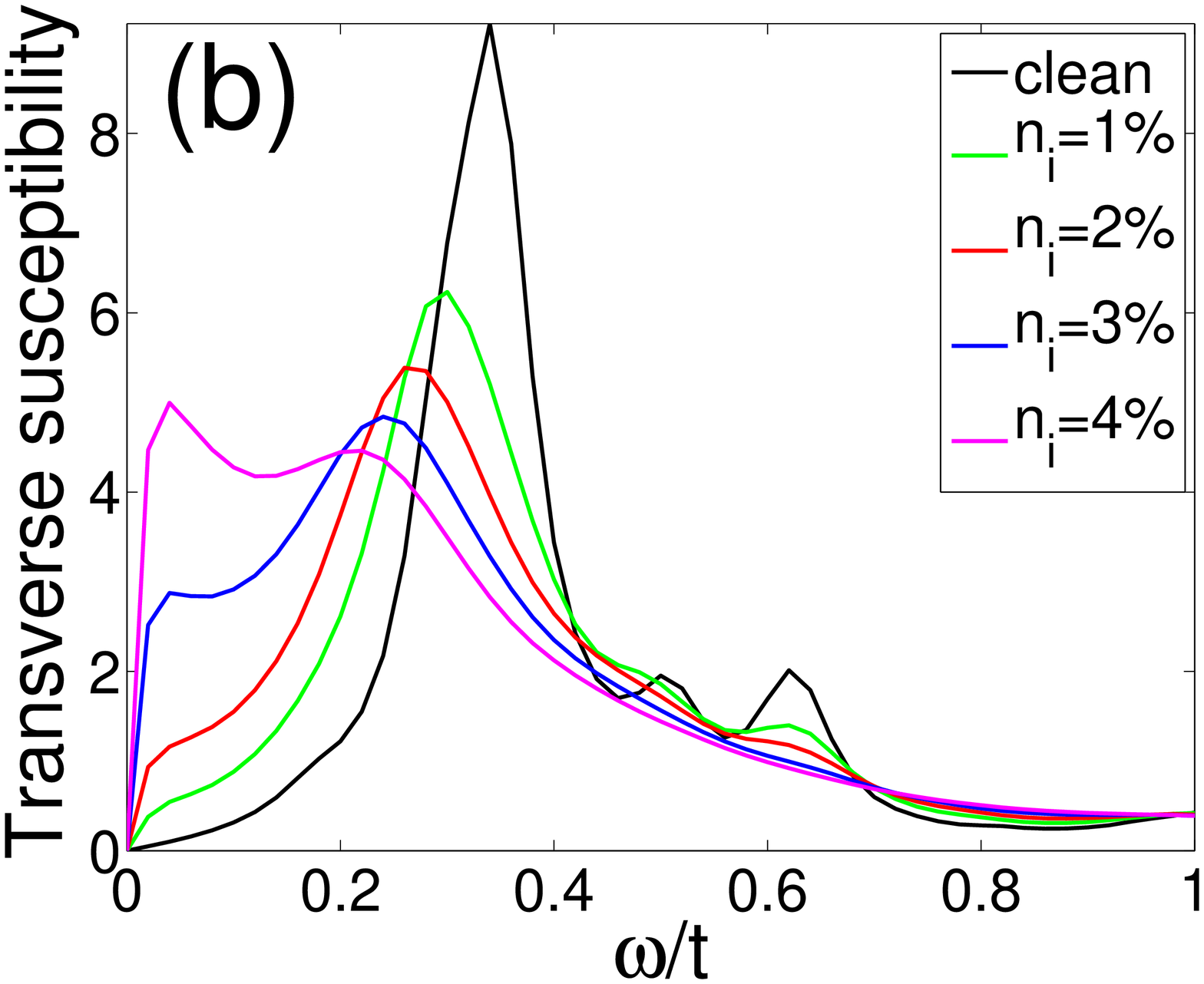}
\end{minipage}
\\
\begin{minipage}{.49\columnwidth}
\includegraphics[clip=true,height=0.8\columnwidth,width=0.98\columnwidth]{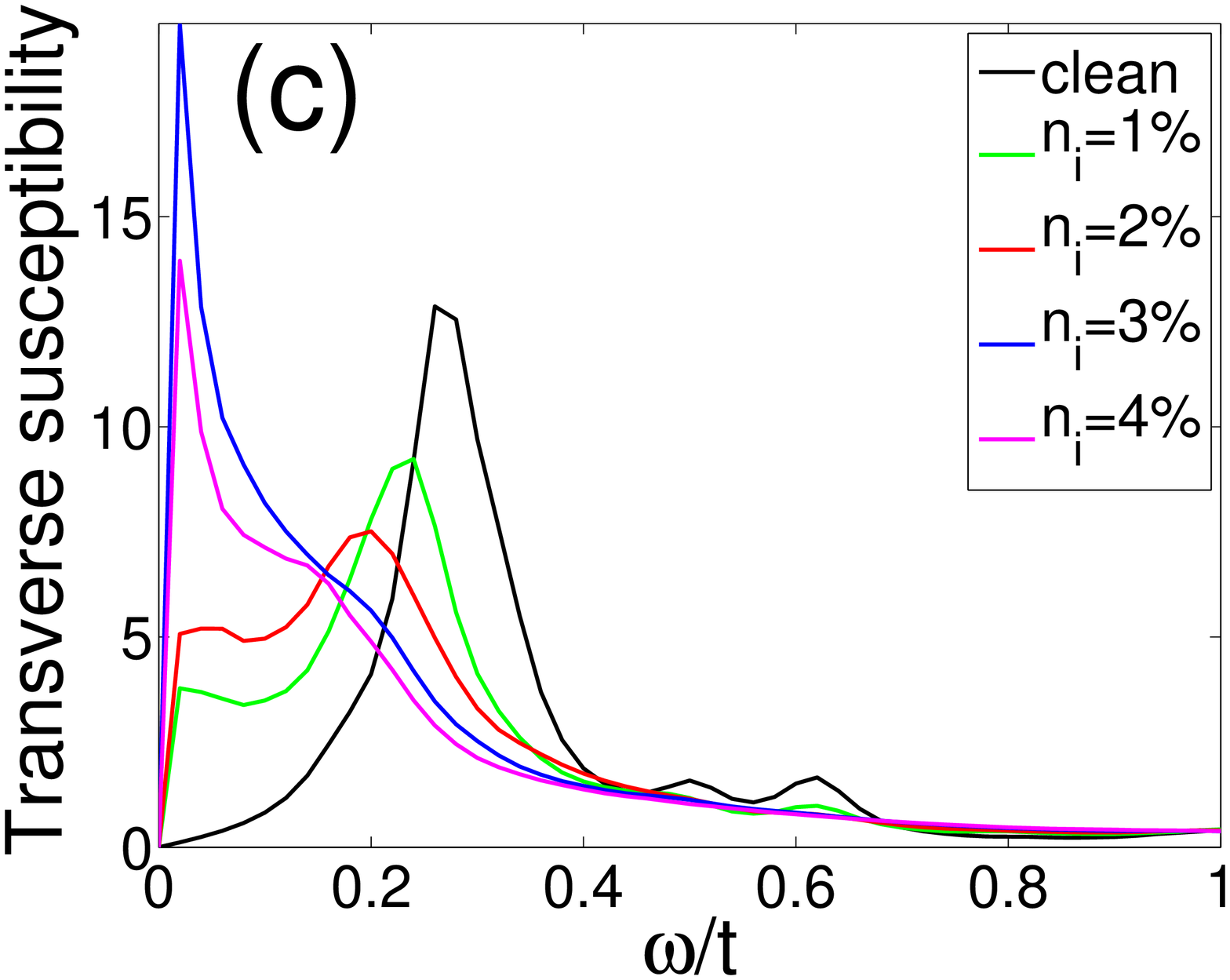}
\end{minipage}
\begin{minipage}{.49\columnwidth}
\includegraphics[clip=true,height=0.8\columnwidth,width=0.98\columnwidth]{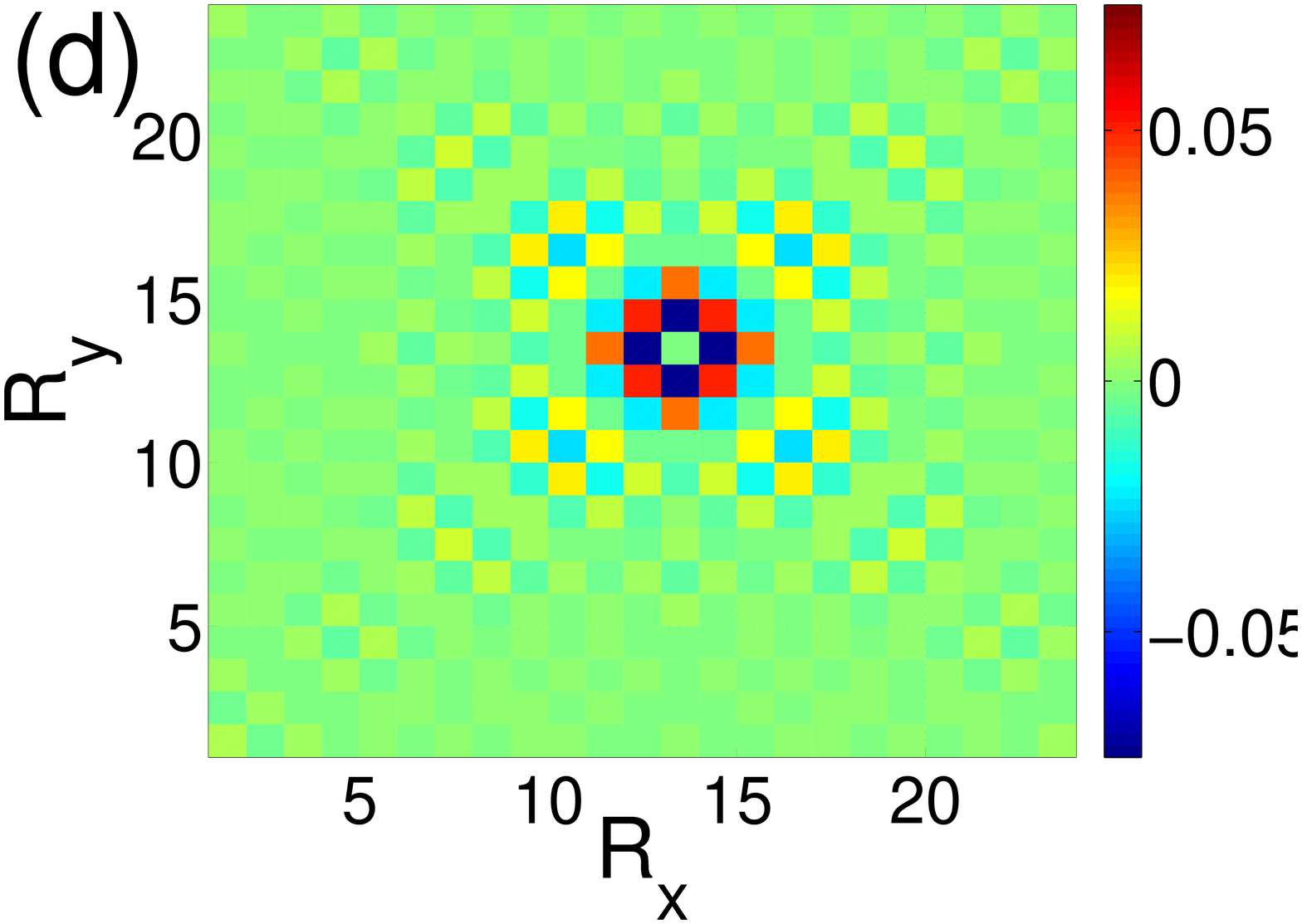}
\end{minipage}
\\
\begin{minipage}{.49\columnwidth}
\includegraphics[clip=true,height=0.8\columnwidth,width=0.98\columnwidth]{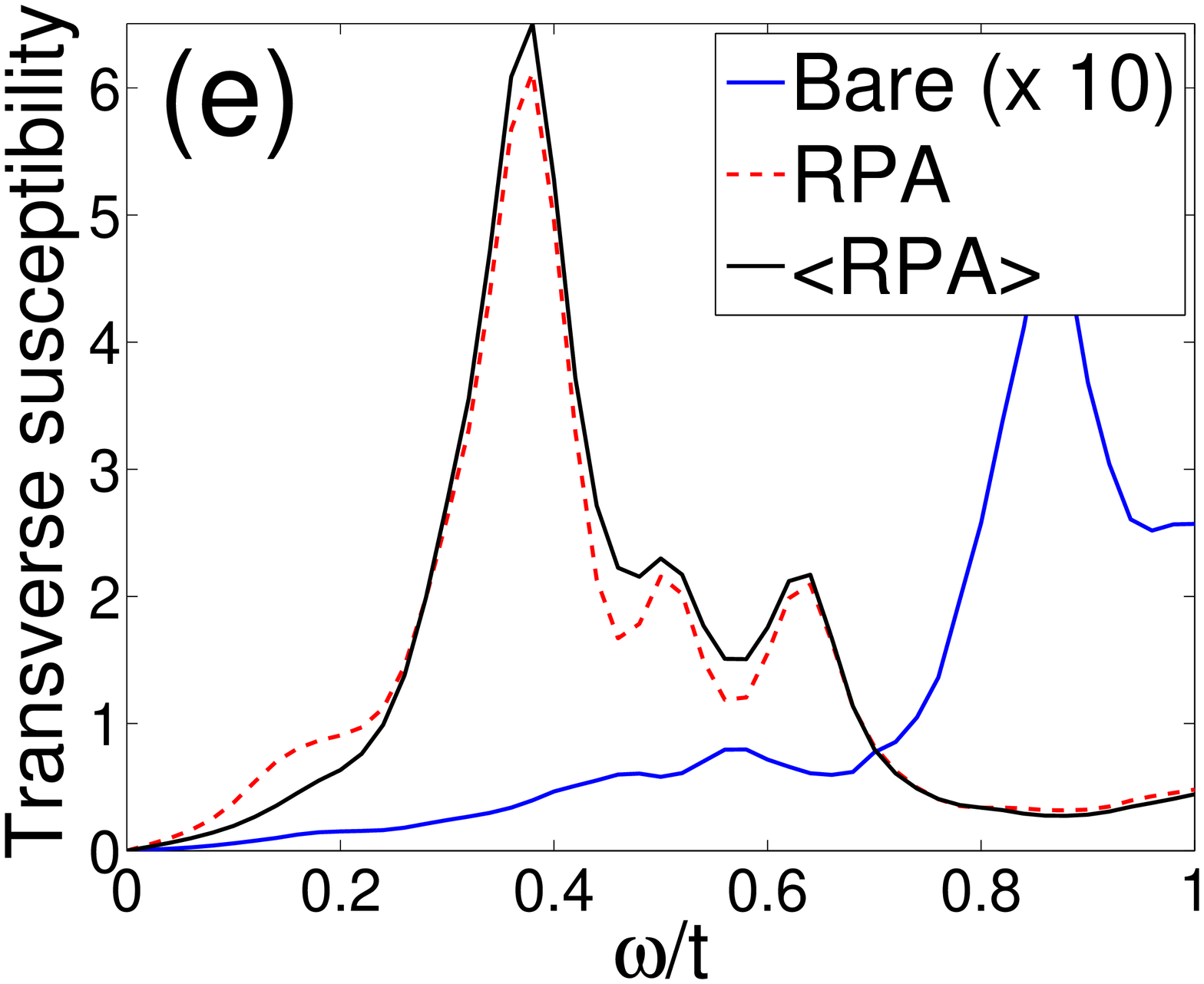}
\end{minipage}
\begin{minipage}{.49\columnwidth}
\includegraphics[clip=true,height=0.8\columnwidth,width=0.98\columnwidth]{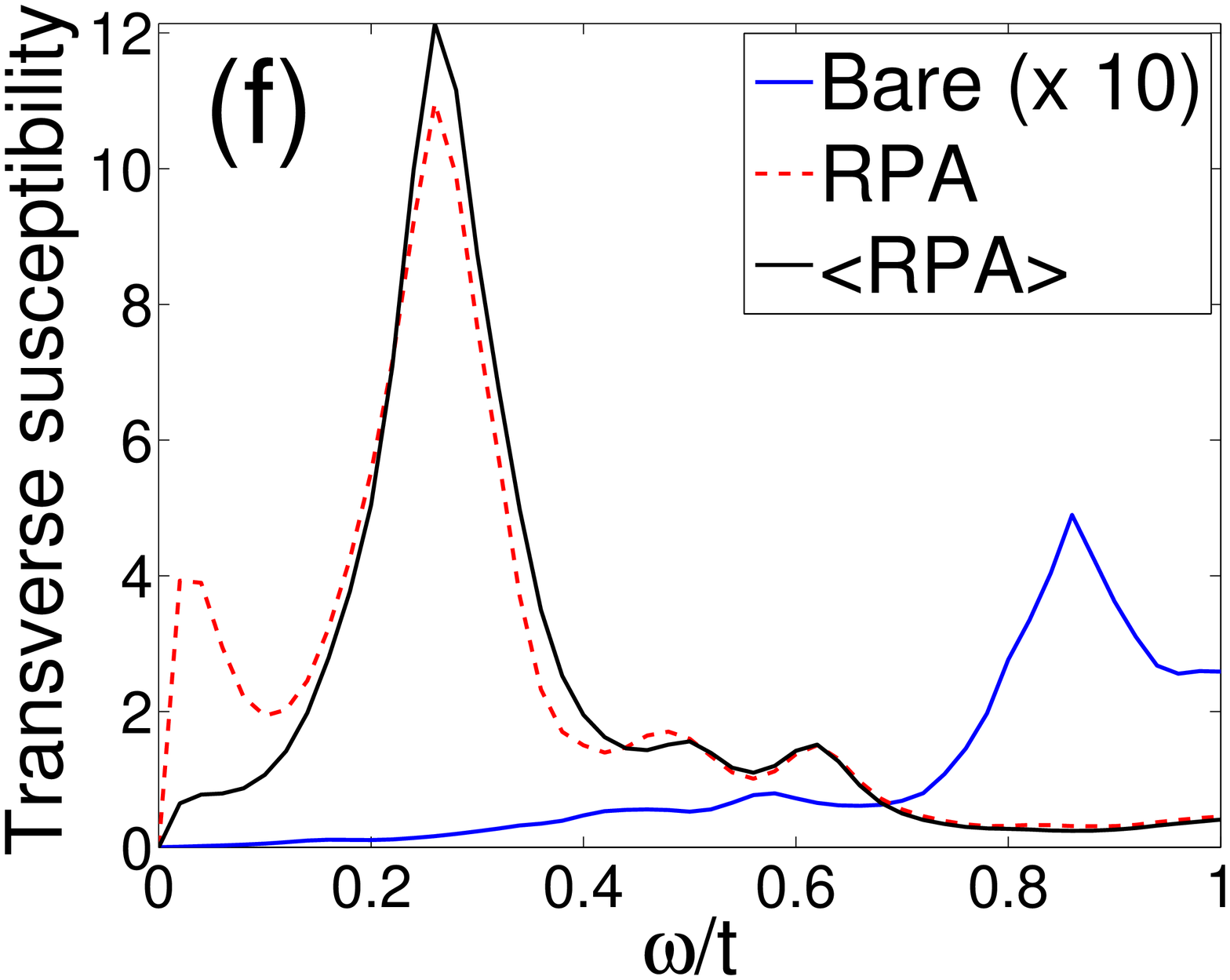}
\end{minipage}
\caption{(a-c) $\chi(q_{IC},\omega)$ as a function of $\omega$ for different $n_i$ and $U=2.5t$ (a), $U=2.6t$ (b), and $U=2.7t$ (c). (d) shows the real-space magnetization induced by a single nonmagnetic impurity at R=(12,12) for $U=2.7t$. (e,f) display the susceptibility at the IC wavevector $q_{IC}=(\frac{3}{4},1)\pi$ in the bare ($U=0$) case (blue solid curve), and the full local susceptibility at $R=(11,12)$ (red dashed line) with $U=2.5t<U_{c1}$ (e) and  $U=2.7t>U_{c1}$ (f). The black solid lines in (e,f) show the spatially averaged full RPA susceptibility $\chi(q_{IC},\omega)$.} \label{fig3}
\end{figure}

We now demonstrate how disorder masks the clear distinction between the susceptibilities in Fig. \ref{fig1}(a,c). In fact by increasing the concentration of nonmagnetic scatterers one may tune the spin response from that in Fig. \ref{fig1}(a) to \ref{fig1}(c). This is evident from Fig. \ref{fig2} showing the evolution of the spectrum in Fig. \ref{fig1}(a) as a function of increased disorder concentrations $n_i$. One clearly sees the impurity-induced low-energy spectral weight entering the legs of the lower part of the hourglass, as well as an overall shifting down and broadening of the dispersion branch. The shift is seen more clearly in Fig. \ref{fig3}(a-c), which show $\omega$-cuts at $q_{IC}=(\frac{3}{4},1)\pi$ for various values of $U$ and $n_i$. At lower $U$, impurity concentrations of just a few percent may strongly enhance the low-energy weight and generate an in-gap mode [Fig. \ref{fig3}(a,b)], in agreement with neutron scattering on Zn-substituted samples, and in applied fields\cite{sidis96,HKimura:2003,suchaneck,chang09,andersen09}.
For sufficiently large disorder concentrations $n_i$ and repulsion $U$, the static component dominates $\chi(q_{IC},\omega)$, as shown in Fig. \ref{fig3}(c).

The results shown in Figs. \ref{fig2}-\ref{fig3} arise from the impurity-induced magnetism as can be directly verified by the spatial information contained in $\chi(q,R,\omega)$. Figures \ref{fig3}(e,f) compare the dynamical susceptibility  of the clean system $\chi(q_{IC},\omega)$  [Fig. \ref{fig3}(e)] to the same quantity measured at the nearest neighbor site of   a single nonmagnetic impurity [Fig. \ref{fig3}(f)]. The local antiferromagnetic (AF) droplet induced by the impurity is shown in Fig. \ref{fig3}(d). From Fig. \ref{fig3}(f) it is evident that the NN-sites to the impurity exhibit a low-energy peak which is responsible for the low-energy weight in the spatially averaged $\chi(q_{IC},\omega)$.

Figures \ref{fig2}-\ref{fig3} apply to the so-called droplet phase where the system spontaneously generate impurity-induced local AF puddles as shown on the right panels of Fig. \ref{fig2}. Note that the susceptibility presented here is for a few percent strong scatterers representing e.g. Zn but that the overall phenomenology may also apply to intrinsically disordered cuprates like LSCO and BSCCO\cite{andersen07,schmid2}.

\begin{figure}[t]
\begin{minipage}{.49\columnwidth}
\includegraphics[clip=true,height=0.8\columnwidth,width=0.98\columnwidth]{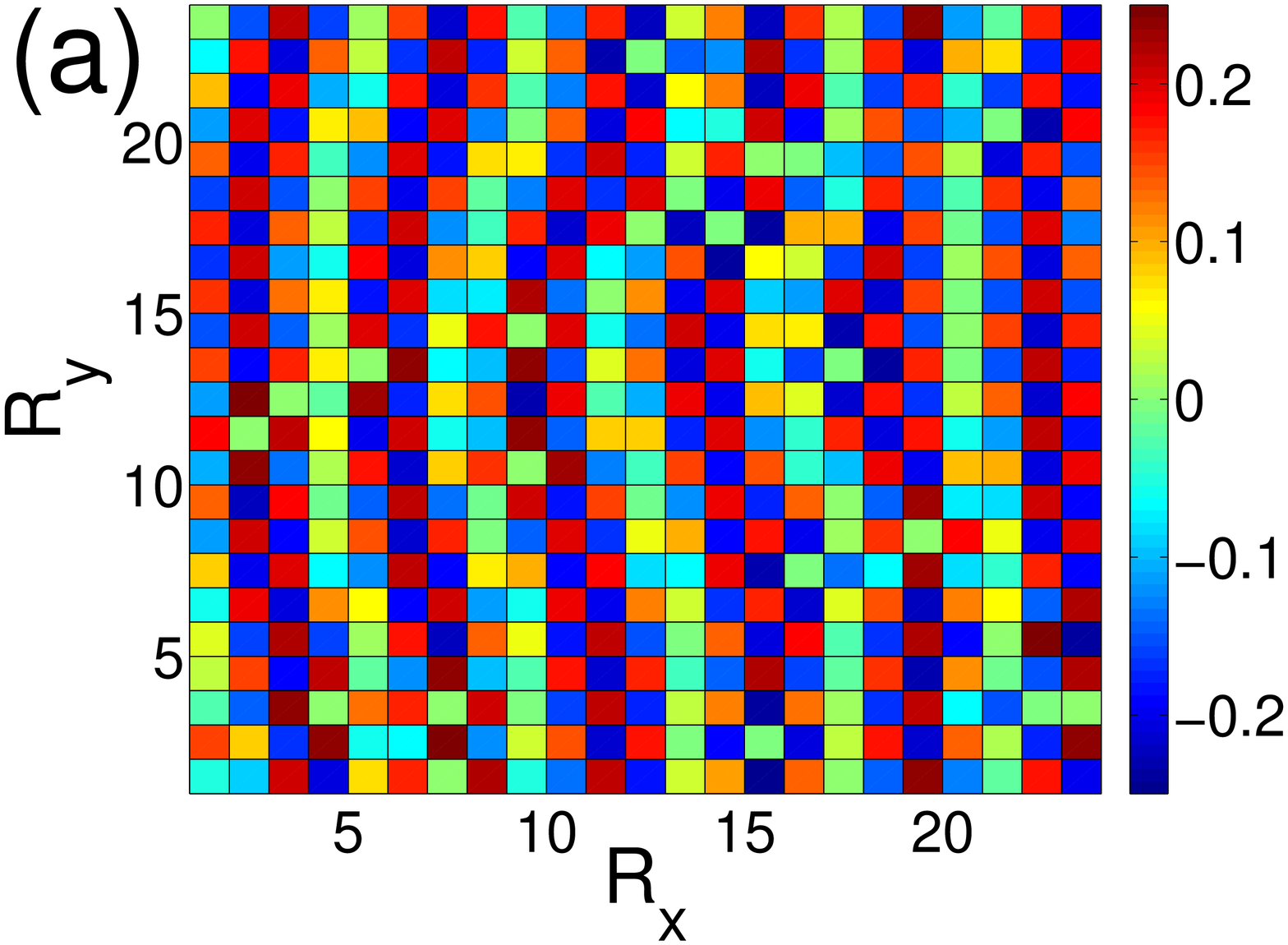}
\end{minipage}
\begin{minipage}{.49\columnwidth}
\includegraphics[clip=true,height=0.8\columnwidth,width=0.98\columnwidth]{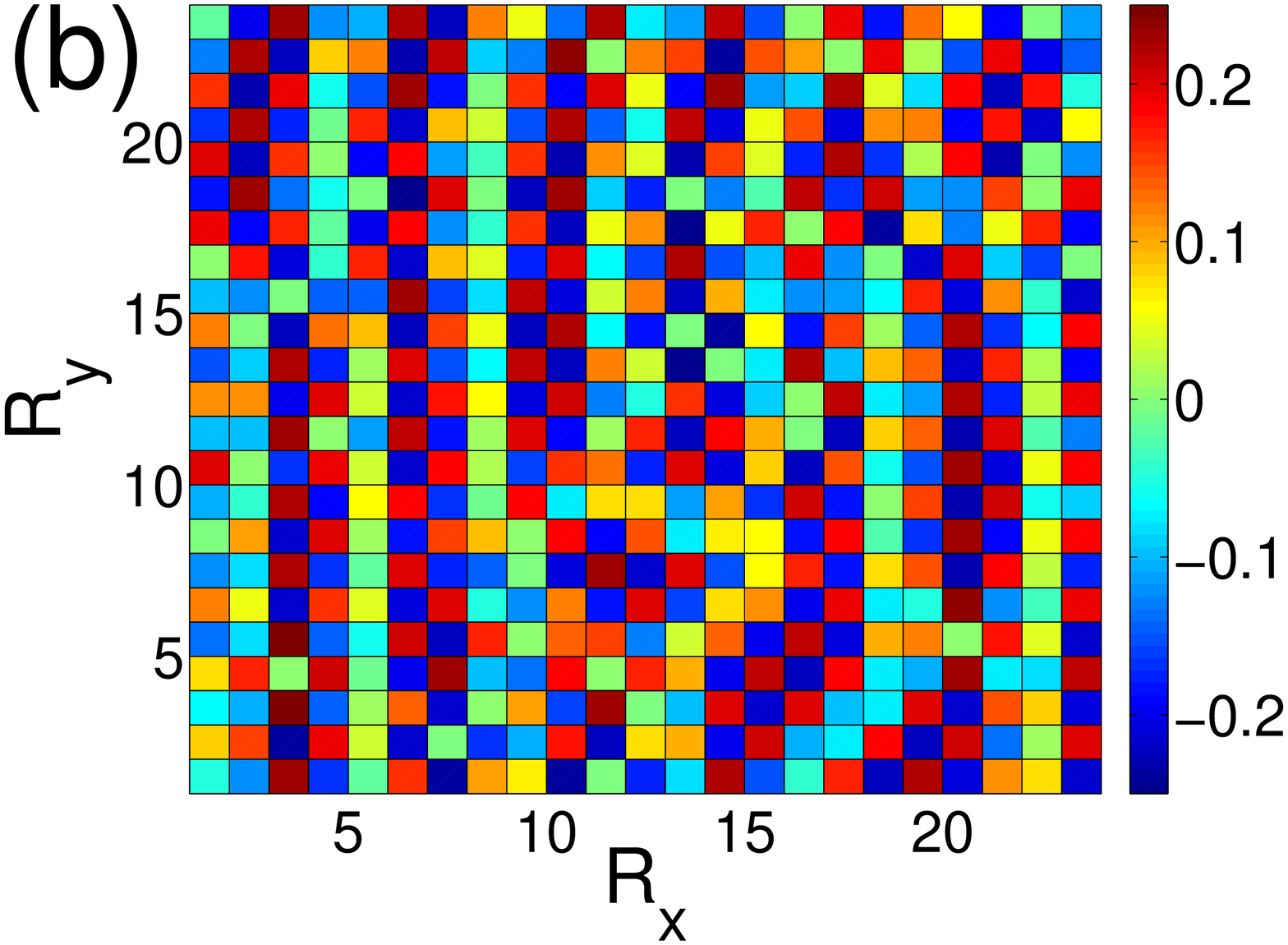}
\end{minipage}
\\
\begin{minipage}{.49\columnwidth}
\includegraphics[clip=true,height=0.8\columnwidth,width=0.98\columnwidth]{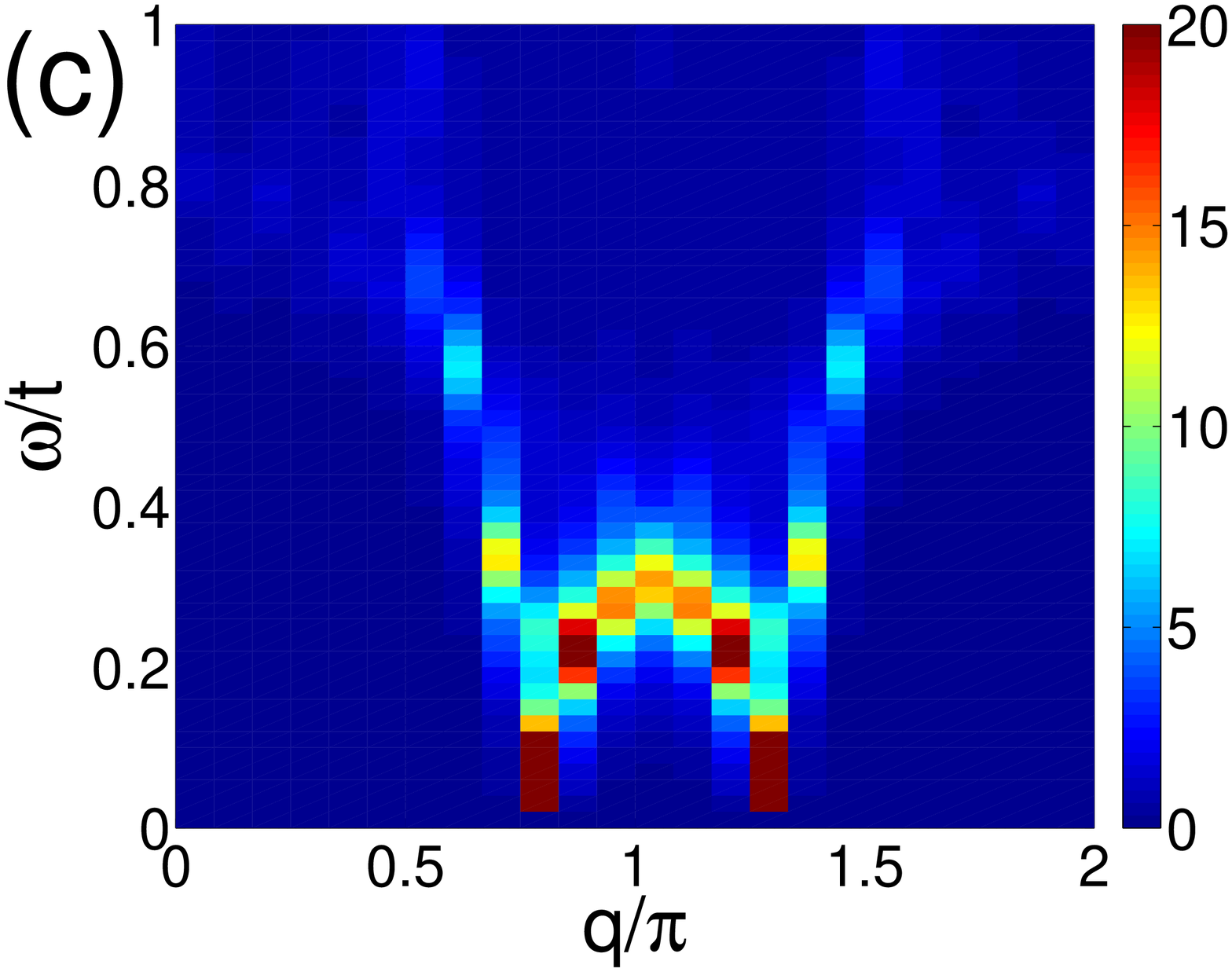}
\end{minipage}
\begin{minipage}{.49\columnwidth}
\includegraphics[clip=true,height=0.8\columnwidth,width=0.98\columnwidth]{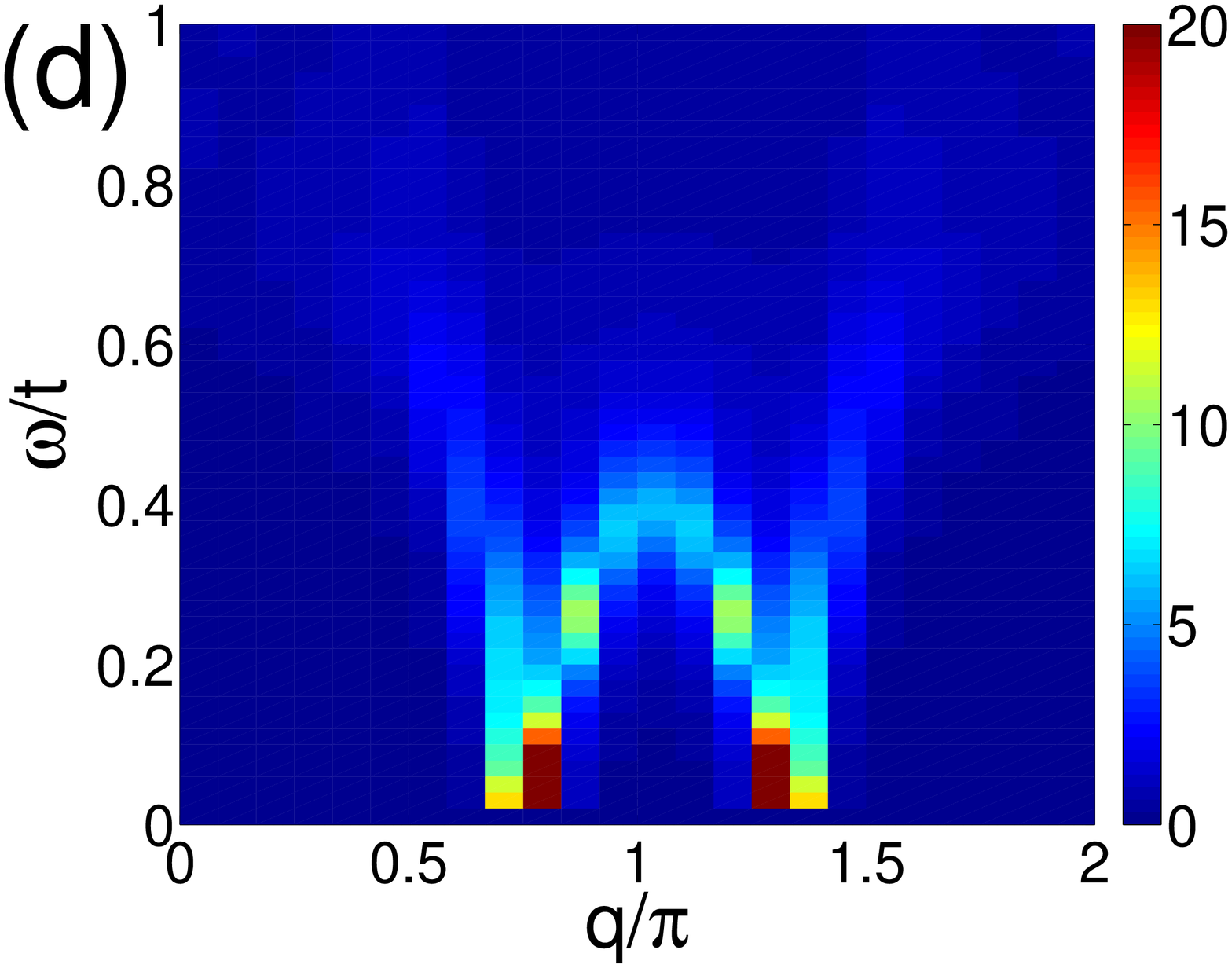}
\end{minipage}
\\
\begin{minipage}{.49\columnwidth}
\includegraphics[clip=true,height=0.8\columnwidth,width=0.98\columnwidth]{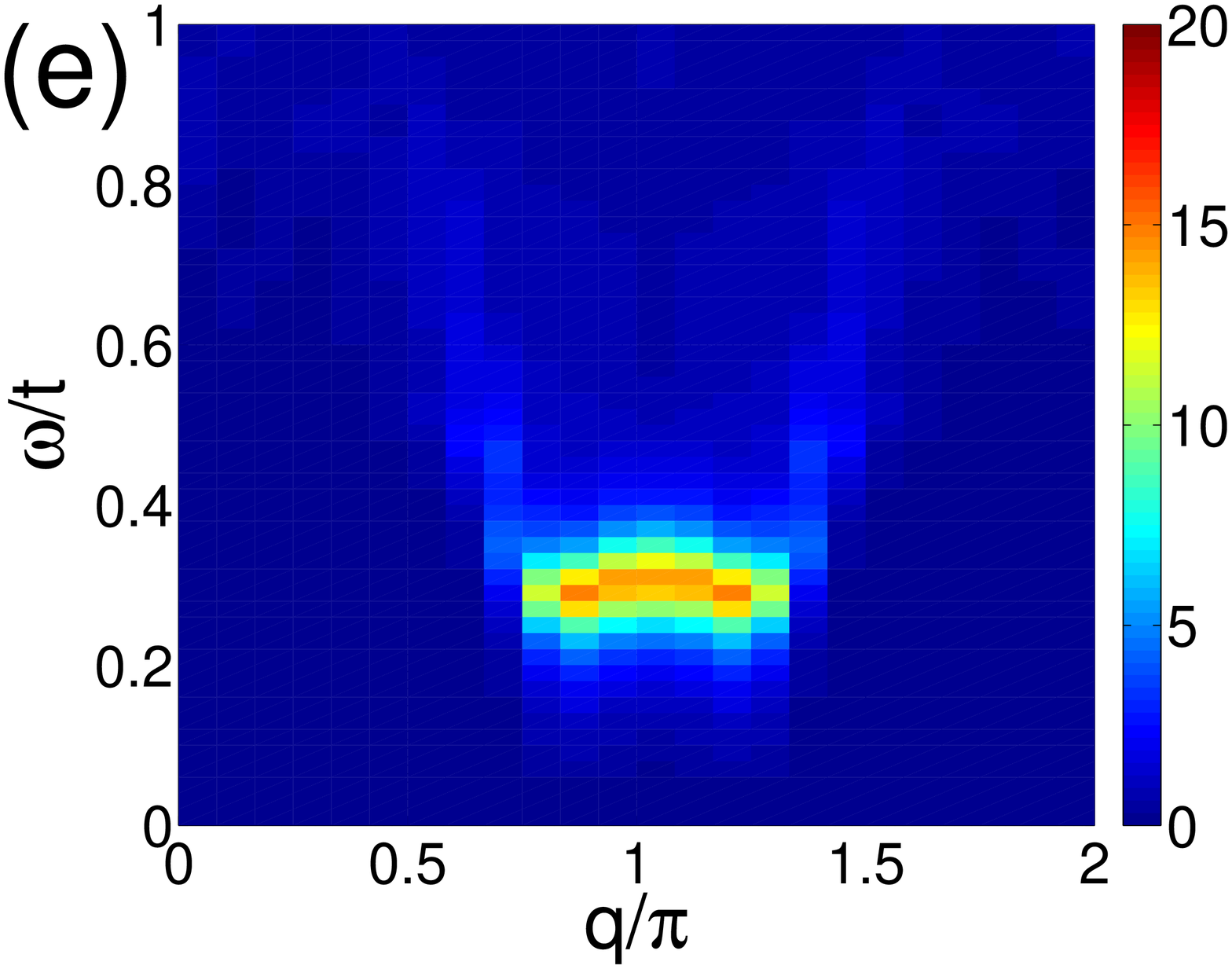}
\end{minipage}
\begin{minipage}{.49\columnwidth}
\includegraphics[clip=true,height=0.8\columnwidth,width=0.98\columnwidth]{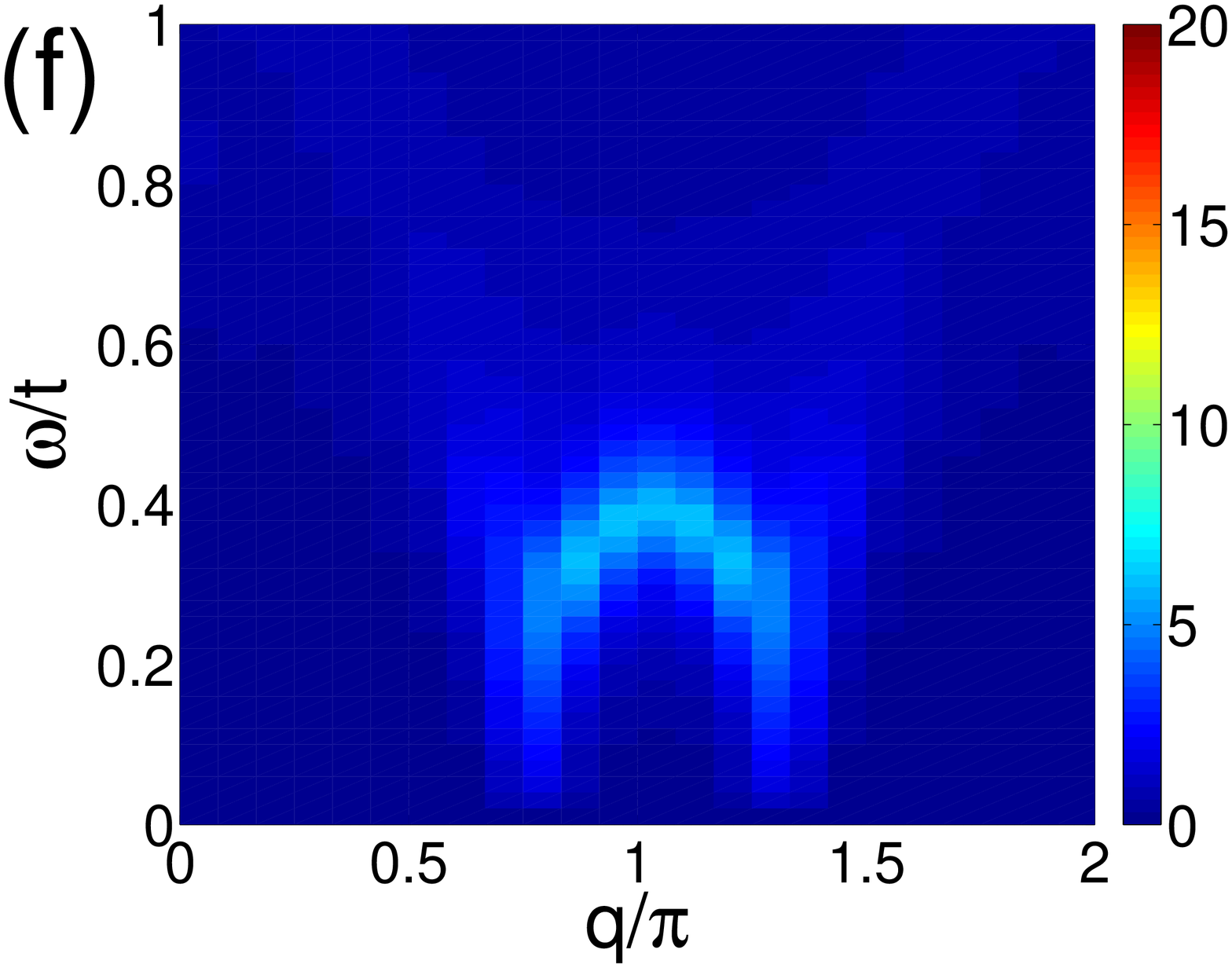}
\end{minipage}
\caption{(a,b) Real-space magnetization with 2\% (a) and 4\% (b) disorder added to a stripe ordered system similar to Fig. \ref{fig1}(d). (c-f) show the associated susceptibility transverse (c,d) and along (e,f) the stripe direction for the same impurity concentrations, i.e. 2\% disorder in (c,e) and 4\% in (d,f).} \label{fig4}
\end{figure}

In the remainder we turn to the limit where $U>U_{c2}$ and compare the results obtained above to a disordered stripe phase\cite{kaul,alvarez,robertson,delmaestro, andersen07,atkinson,andersen08,schmid2}.  Figure \ref{fig4}(a,b) show how the magnetization from Fig. \ref{fig1}(d) evolves into a nematic phase when disorder destroys the translational invariance. For a discussion of the disorder effects on the spin spectrum it is convenient to look separately at the dispersion perpendicular [Fig. \ref{fig4}(c,d)] and parallel  [Fig. \ref{fig4}(e,f)] to the stripe direction. In the limit of low disorder [Fig. \ref{fig4}(a,c,e)] the spin fluctuations are qualitatively similar to the clean stripe case from Fig. \ref{fig1}(c) (which shows the directional-averaged susceptibility). When sufficient disorder is added, the stripes break up and meander strongly, which is revealed in the spin response as a significant broadening of the high-energy part of the hourglass including the spin resonance at $(\pi,\pi)$, and a shifting-down of spectral weight in the response parallel to the stripe direction as seen from Fig. \ref{fig4}(f). A natural question arising from the similarity between the spin spectrum in the droplet phase [see e.g. Fig. \ref{fig2}(e)] and the disordered stripe phase in Fig. \ref{fig4} is how to experimentally distinguish between them. Since the droplet phase preserves the $C_4$ symmetry of the underlying lattice whereas the stripe phase breaks $C_4$ to $C_2$ which remains robust even with significant disorder, we find that $q_x-q_y$ anisotropy in the neutron response is the best way to answer this question. Experiments on detwinned underdoped YBCO samples have recently reported such anisotropy in the low-energy intrinsic\cite{hinkov} and Zn-induced spectral weight\cite{suchaneck}. Hopefully future experiments may reveal whether similar results apply to e.g detwinned LSCO crystals. Careful studies of any potential anisotropy in the inelastic signal may also reveal to what extent the droplet phase can be viewed as a precursor of the stripe phase\cite{schmid2}.

{\it Conclusions.} We have studied disorder effects on the hourglass spin fluctuation spectrum relevant for the cuprates.
We have focussed on two regimes of electronic repulsion $U$ leading to a magnetic droplet phase for low $U$ and a disordered stripe
phase at larger $U$. Within this scenario, it is likely that the underdoped state is 
 well described by our results for larger $U$, since we expect that the effective interaction scale $U/t$
 increases as one approaches the Mott limit.
 We have also shown how neutron studies of detwinned samples may be used to distinguish between them despite their
surprisingly similar fluctuation spectrum in as-grown samples.

B.M.A. acknowledges support from The Danish Council for Independent Research $|$ Natural Sciences. S.G. 
acknowledges support from the DFG through TRR 80.
P.J.H. received partial support from the U.S.  Dept. of Energy  under grant
DE-FG02-05ER46236.

\end{document}